\documentclass[11pt]{article}

\usepackage[final]{acl}
\usepackage{multirow}
\usepackage{amsmath}
\usepackage{times}
\usepackage{latexsym}
\usepackage{amssymb}
\usepackage{algorithm}
\usepackage{algpseudocode}
\usepackage[T1]{fontenc}
\usepackage[utf8]{inputenc}

\usepackage{microtype}

\usepackage{inconsolata}

\usepackage{graphicx}
\usepackage{subfigure}
\usepackage{utfsym}
\title{BotUmc: An Uncertainty-Aware Twitter Bot Detection with Multi-view Causal Inference }

\author{Tao Yang \and Yang Hu \and Feihong Lu \and Ziwei Zhang \and Qingyun Sun \and Jianxin Li\\
School of Computer Science and Engineering, \\ BDBC, Beihang University, Beijing, China }

\begin{document}
\maketitle
\begin{abstract}
Social bots have become widely known by users of social platforms. To prevent social bots from spreading harmful speech, many novel bot detections are proposed. However, with the evolution of social bots, detection methods struggle to give high-confidence answers for samples. This motivates us to quantify the uncertainty of the outputs, informing the confidence of the results. Therefore, we propose an uncertainty-aware bot detection method to inform the confidence and use the uncertainty score to pick a high-confidence decision from multiple views of a social network under different environments. Specifically, our proposed BotUmc uses LLM to extract information from tweets. Then, we construct a graph based on the extracted information, the original user information, and the user relationship and generate multiple views of the graph by causal interference. Lastly, an uncertainty loss is used to force the model to quantify the uncertainty of results and select the result with low uncertainty in one view as the final decision. Extensive experiments show the superiority of our method.
%Therefore, we propose an uncertainty-aware bot detection to inform the confidence and use the uncertainty score to pick a high-confidence decision from multiple views of a social network under different environments. Also, a modified uncertainty loss is designed to fix false determinations with substantial evidence, with the aim of avoiding misclassification due to disguised features of bots. Finally, extensive experiments demonstrate the superiority of our method.
%社交网络上的机器人检测已经成为一项维护社交平台的重要任务，%
\end{abstract}

\section{Introduction}
Social media provides convenience for communication and information acquisition in people's daily lives and allows for spreading misinformation \cite{starbird2019disinformation,zannettou2019disinformation}, election interference~\cite{ferrara2017disinformation}, and terrorist propaganda~\cite{chatfield2015tweeting}, to which social bots can be credited. To reduce the risk posed by bots, extensive research efforts~\cite{feng2021botrgcn,feng2022heterogeneity,liu2023botmoe} have investigated ways to distinguish bots from humans.%, using the Twitter platform as an example.

	\begin{figure}[t]
		\vskip 0.2in
		\centering
		\subfigure[Comparison of the previous methods and ours]{
			\includegraphics[width=0.85\linewidth]{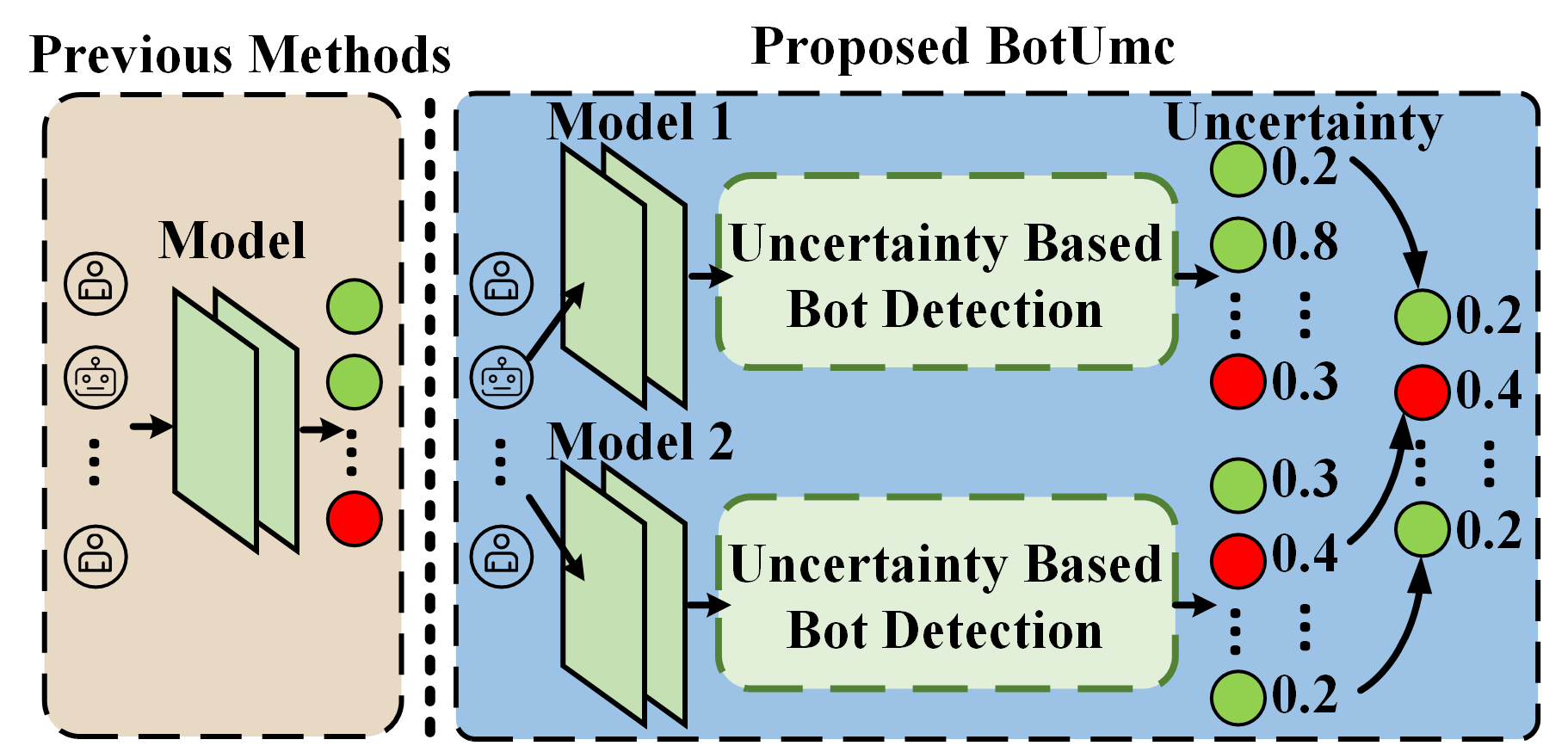}
			\label{fig:comp_model}
		}
        
		\subfigure[Relationship between uncertainty and performance]{
			\includegraphics[width=0.9\linewidth]{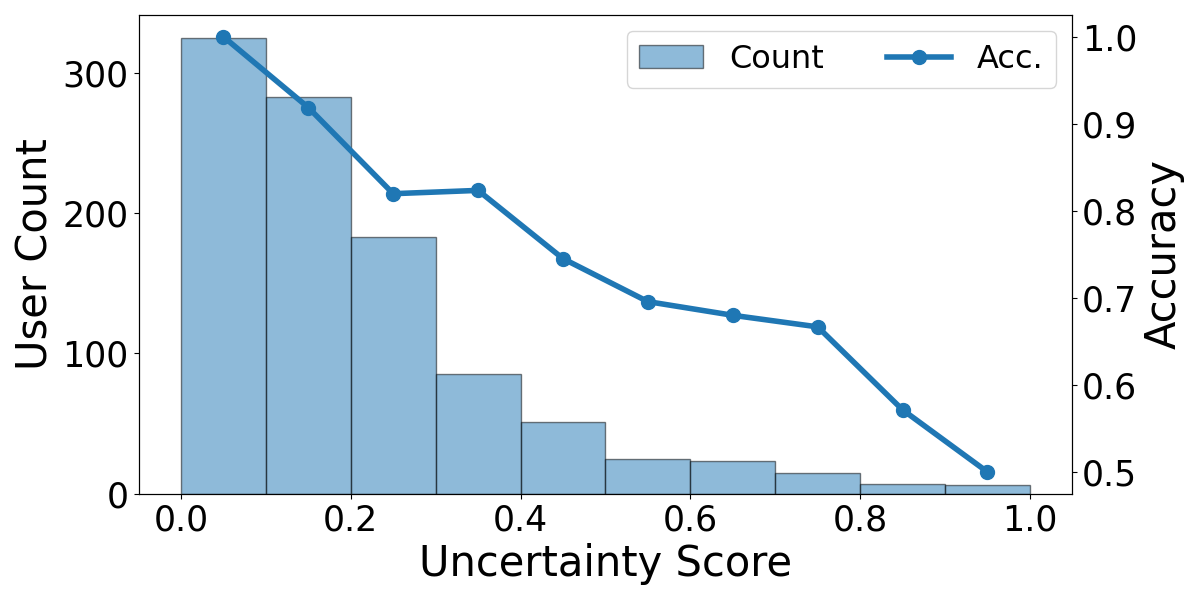}
			\label{fig:comp_result}
		}
		\caption{ (a) Comparison between previous Twitter bot detection and our uncertainty-aware bot detection. (b) Relationship between the accuracy of bot detection and the uncertainty of results. The bins indicate the count of users whose results are within a certain range of uncertainty. The lines show the corresponding performance. }
		\label{fig:comp}
		\vskip -0.2in
	\end{figure}
  
Existing social bot detection methods can be divided into three categories: feature-based methods, text-based methods, and graph-based methods. Feature-based methods \cite{kudugunta2018deep,hayawi2022deeprobot} perform feature engineering on tweets and metadata to extract key information and detect social bots by traditional classification algorithms. The success of feature-based detectors forced bot developers to put in place sophisticated countermeasures by making bots with advanced features. With the evolution of bots, feature-based methods become increasingly difficult to cope with the bots with advanced features \cite{cresci2020decade,cresci2017paradigm}. Thanks to the emergence of deep learning, text-based methods \cite{wei2019twitter,feng2021satar} are presented to detect bots by natural language processing technology. As a result, bot developers steal text from real users to deceive text-based methods. To mitigate this issue, graph-based methods \cite{feng2021botrgcn,liu2023botmoe} introduce Graph Neural Networks (GNNs) to utilize more comprehensive user information, such as topological structure information.
  
% Existing social bot detection methods can be divided into three categories: feature-based methods, text-based methods, and graph-based methods. Feature-based methods use user information for feature engineering and apply traditional classification algorithms to detect bots. Text-based methods use natural language processing technology to process the text content of user tweets and user description texts, and detect Twitter bots through malicious text content. Graph-based methods interpret users in social networks as nodes and relationships as edges, and use graph neural networks (GNNs) \cite{scarselli2008graph} to iteratively aggregate representations from neighbors to learn node representations for bot detection. Recent studies\cite{liu2023botmoe, feng2021botrgcn, feng2022heterogeneity} have shown that 

Although the existing graph-based methods have achieved great performance in social bot detection, the low confidence of some predictions due to insufficient information has not received attention. Therefore, this work tries to design an uncertainty quantification for bot detection, achieving uncertainty-aware bot detection, as shown in Figure \ref{fig:comp} (a). The results in Figure \ref{fig:comp} (b) show that our method achieves a spurious association between performance and uncertainty, demonstrating that the more certain the result, the more likely it is to be correct. Thanks to uncertainty quantification, we can pick the more reliable human-bot association from multiple views, improving the results of bot detection.

Our contributions can be summarized as follows:
\begin{itemize}
\item We propose an uncertainty-aware Twitter bot detection framework, dubbed BotUmc, that can say "I am not sure about the result". Extensive experiments demonstrate that BotUmc achieves great performance on three datasets.

\item An uncertainty quantification module for bot detection is proposed to measure the reliability of outputs so that the more reliable output can be chosen as the final decision.

\item We introduce the causal intervention to construct multiple views of graphs by simulating different environments and thereby find high-confidence features.

\end{itemize}

\section{Related Work}
\textbf{Twitter Bot Detection.} 
Existing Twitter bot detection methods can generally be divided into three categories: feature-based, text-based, and graph-based methods. 

Feature-based methods perform feature engineering based on user metadata and then combine it with traditional classification algorithms. \citet {kudugunta2018deep} used a deep neural network based on a contextual LSTM architecture to extract user metadata features to detect bots; \citet{miller2014twitter} used tweet content features for detection; \citet {hayawi2022deeprobot} used a hybrid architecture of LSTM units and dense layers to process mixed features. However, as bots evolve, they tamper with features to evade detection \cite {cresci2020decade}.

Text-based methods use natural language processing technology to detect based on tweets and user descriptions. \citet{lei2022bic} pointed out that there would be semantic inconsistencies between tweets posted by robots and those posted by humans; \citet {wei2019twitter} used BiLSTM in RNN for detection; \citet {feng2021satar} proposed a self-supervised learning framework to jointly encode multiple types of information for detection; \citet {dukic2020you} used the BERT-BASE model to encode tweets. However, advanced bots can evade such detection that only analyzes the content of tweets by copying real user texts.

Graph-based methods attempt to construct social networks into a graph structure, with users as nodes and relationships as edges, and perform detection through GNN. \citet {dehghan2023detecting} used a structural embedding algorithm, \citet {pham2022bot2vec} improved the Node2Vec algorithm, \citet {magelinski2020graph} used latent local features of graphs, and \citet {liu2023botmoe} proposed community-aware modality-specific expert hybrid detection. GNN constructs heterogeneous graph aggregation representation to achieve advanced detection performance, but bots use graph strategies to construct false associations to evade detection. %However, existing bot detection methods ignore the confidence of the model output and fail to provide reliable predictions.

\textbf{Uncertainty Estimation.}  The methods of generating uncertainty can be divided into two main categories. %In the field of deep learning, as the number of model parameters continues to rise, how to control these models and ensure the reliability of their outputs has become the focus of researchers. At present,
The first category is to force the model to learn the uncertainty of the output directly. Examples include evidence learning \cite{sensoy2018evidential,amini2020deep}, Bayesian neural network \cite{mackay1992practical}, deep integration \cite{lakshminarayanan2017simple}, random weight average (SWAG) \cite{maddox2019simple} and Monte Carlo Dropout \cite{gal2016dropout}. However, such methods often require multiple iterations and estimates to optimize the entire model parameters, which makes them more suitable for models with fewer parameters.
The second category of methods uses techniques such as transfer learning \cite{kandemir2015asymmetric} or distillation learning \cite{fathullah2022self}. Through these methods, models with many parameters can effectively learn output uncertainty by optimizing only specific layers instead of the entire model parameter set. This not only improves the efficiency of the optimization process but also helps maintain reasonable uncertainty in the model output. In addition, VBLL \cite{harrison2024variational} proposed a deterministic variational formula for training the last layer of Bayesian neural networks, which greatly improved the efficiency of uncertainty estimation.%Because existing uncertainty estimation methods lack the ability to learn difficult samples, they are not directly applicable to social robot detection tasks.

\begin{figure*}[h]
    \setlength{\belowcaptionskip}{-0.3cm}
    \centering
    \includegraphics[width=0.95\linewidth]{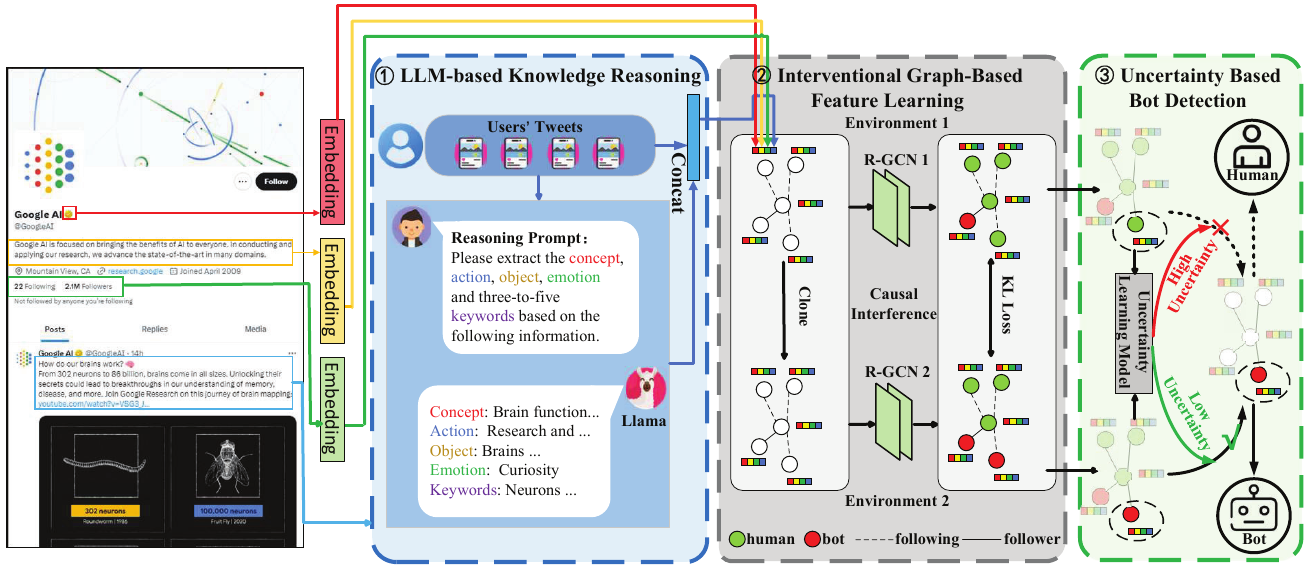}
    \caption{The overview of our proposed BotUmc. It jointly utilizes multiple types of user information: Text, Metadata, and Topology information to detect bots. Twitter users’ tweets are first processed by the LLMs module, then encoded with other user information, and then processed by the causal interference module. Finally, the uncertainty module is used to integrate Twitter users under multiple views to classify them.}
    \label{fig:figure2}
\end{figure*}
\section{Problem Formulation}
In this section, we define the social bot detection task using multiple types of user information. 

Given a user, the corresponding information is represented by $x_i \in X$, where $x_i$ contains the following components:
\textbf{Text information:} includes the user description and the tweets posted by the user, which are represented by $T$.
% user description $B_i$ and tweets $S_i = \{ S_{i,j} \}_{j=1}^{T_i}$, where $T_i$ is the number of tweets.
\textbf{Metadata information:} numerical data $N_i$ (e.g., number of followers, likes) and Boolean data $C_i$ (e.g., verification status). Combining with the information of relationships between users, we construct a heterogeneous graph $G = G(X, E,  R_e, Y)$, where $E$ is the edge set, 
% $\varphi : E \to R_e$ maps edges to relationships, 
$R_e$ is the relationship type set, and $Y$ is the label of users. 
% The neighbors of $u_i$ are derived from this graph and represented as $N_i = \{ n_{i,j} \}_{j=1}^{J_i}$, where $J_i$ is the number of neighbors.
Therefore, we aim to construct a detection function $f(G)$ which can output predict values $\hat{Y} \in \{0, 1\}$ and uncertainty scores $U \in \{0, 1\}$.
% , where 0 represents a human and 1 represents a bot.

\section{Method}
A large number of bots on social platforms mimic real user behaviors to deceive humans and bot detection. To tackle this challenge, as shown in Figure \ref{fig:figure2}, we propose an uncertainty-aware Twitter bot detection framework, named BotUmc, which proceeds to: 
\textbf{1)} The LLM-based knowledge reasoning module employs large language models (LLMs) to extract key insights from users' tweet texts; 
\textbf{2)} The Interventional Graph-Based Feature Learning module constructs a heterogeneous social graph and generates multiple views of the graph through the multi-environment causal intervention. This helps to detect spurious associations between humans and bots.
\textbf{3)} The uncertainty-based bot detection module evaluates node classification reliability through uncertainty quantification and selects more reliable output as a prediction from multiple views of the graph to detect disguised bots accurately.

\subsection{LLM-based Knowledge Reasoning}
Human tweets are often implied rather than explicit, and social bots often disguise their intentions by inserting elements into the content of their tweets. This makes the “noise” common in social media tweets greatly amplify the complexity of extracting information from tweets. The complexity may lead to incorrect information extraction, which in turn causes inaccurate distinctions between social bots and real users. To extract useful information, we propose a key knowledge prompt strategy to guide LLM in extracting concepts, actions, objects, emotions, and keywords from various dimensions of the original tweets. %Additionally, a single tweet on social media is often incomplete, and its content can be unclear. As a result, directly using text for detection

Concretely, we first concatenate multiple tweets from the same user to provide full information. The concatenated information is then fed into the Llama, and the uniquely designed prompt is used to guide the Llama \cite{wang2023chatgpt} to reason about the key information in the user's tweets:
\begin{figure}[h]
    \centering
    \includegraphics[width=0.85\linewidth]{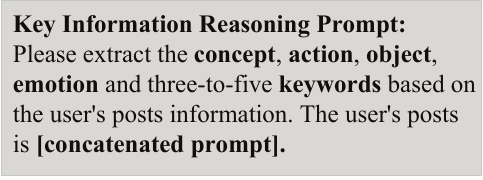}
\end{figure}

Here, we use the LLM to extract the key information from the original tweets:
\begin{equation}\label{equ_Llam}
\small{
\begin{aligned}
t_{\text{key}} = Llama\left( concat\left( \sum_{i=1}^{N_t} t_i \right) \right),
\end{aligned}}
\end{equation}

where $t_{\text{key}}$ is the key knowledge after the LLM, $t_i$ is the user's single tweet, and $N_t$ is the total number of the user's tweets. 

Then, We use the pre-trained RoBERTa\cite{liu2019roberta} model to encode the key information and the connected original tweets and pass it through a layer of MLP to obtain the user's tweet representation $\boldsymbol{v}_{concat}$:

\begin{equation}\label{equ_Roberta}
\small{
\begin{aligned}
\boldsymbol{v}_{concat} &=  MLP \left(RoBERTa \left( \left\{ t_i \right\}_{i=1}^{N_t} ; t_{\text{key}} \right)\right).
\end{aligned}}
\end{equation}

\subsection{Interventional Graph-Based Feature Learning}
\label{sec:feature_learning}
Social bots have evolved to the point where they can effectively mimic human behavior and interaction patterns. They are capable of modeling human behavior and establishing seemingly authentic interactions with human users, thereby creating spurious social associations that further blur the boundaries between humans and bots. This ability to disguise themselves and form deceptive connections reduces the accuracy of bot detection systems, increasing the likelihood of misjudgment. To address the above challenges, we propose an interventional graph-based feature learning approach, which includes 
heterogeneous graph construction, and intervention based graph update.\\

\textbf{Heterogeneous Graph Construction.} To address the challenge of detecting social bots' camouflage, we consider the differences between social bots and humans by leveraging multi-view user information. 
As shown in Figure \ref{fig:figure2}, we integrate text information (user descriptions, user tweets) and metadata information (user numeric attributes and Boolean data) as nodes $X$, and use the interactions between users ("friend" and "follow") as edges $E$ to build a heterogeneous graph. In order to make full use of different types of information, inspired by the feature encoding process in the work \cite{feng2021botrgcn}, we encode the information as follows:

\begin{figure}[t]
    \setlength{\belowcaptionskip}{-0.3cm}
    \centering
    \includegraphics[width=0.8\linewidth]{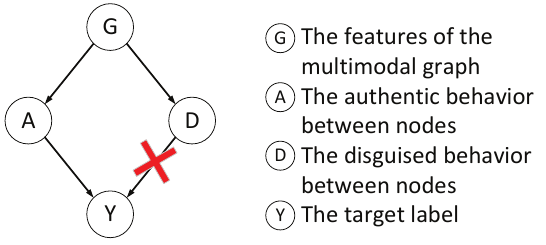}
    \caption{The causal structure.}
    \label{fig:casual}
\end{figure}

For user descriptions (in the yellow box in Figure \ref{fig:figure2}), we use RoBERTa to obtain the description embedding $\boldsymbol{v}_{des}$.
% \begin{equation}\label{equ_user_roberta} \begin{aligned} \bar{T} &= \text{RoBERTa} \left( {T_i} \right), \ \bar{T} &\in \mathbb{R}^{D_s \times 1} \end{aligned} \end{equation}
For metadata, we incorporate both numerical and Boolean attributes derived from the Twitter API. The numerical features (in the green box in Figure \ref{fig:figure2}) are standardized using $z$-score normalization, and their final representation, denoted as $\boldsymbol{v}_{num}$, is obtained through a fully connected layer. The Boolean attributes are encoded via one-hot encoding, followed by concatenation and transformation through a fully connected layer to produce the representation of the user's categorical features (in the red box in Figure \ref{fig:figure2}), denoted as $\boldsymbol{v}_{bl}$.

For each user, we concatenate the encoded representations of the four different features as the user feature vector $\bar{X}$, and ensure that the vector dimensions of each feature are the same:
\begin{equation}\label{equ_all_embedding}
\bar{X} = \left[ \boldsymbol{v}_{des}; \boldsymbol{v}_{concat}; \boldsymbol{v}_{num}; \boldsymbol{v}_{bl} \right].
\end{equation}

Subsequently, we define $\bar{X}$ as the set of nodes and $R_e =\{ \text{followers},\text{likes}\}$ as the set of edges to construct a heterogeneous graph $G_o = G(\bar{X}, E, R_e, Y)$, where $Y$ denotes the label associated with each user.

\textbf{Intervention Based Graph update.} 
We propose a causal view of the training and detection processes underlying the social bot detection task, as shown in Figure \ref{fig:casual}. Here, we use the causal view to examine the causal association between four variables: the features of the heterogeneous graph $G$ (including the structure $E$ on the graph and node features $\bar{X}$), the authentic behavior $C$ between nodes in the graph (including the associations $E_A$ and features $\bar{X}_A$ of real users), the disguised behavior $D$ between nodes in the graph (including the fake associations $E_D$ constructed by bots and the disguised features $\bar{X}_D$ where bots mimic real users), and the target label $Y$. Figure \ref{fig:casual} illustrates the causal association between different variables in the task, which are described in detail as follows.

The heterogeneous graph $G$ consists of authentic behavior $A$ and disguised behavior $D$. $A$ → $Y$ means that the causal variable $A$ is the only endogenous parent node, which is crucial for correctly classifying the entity type; $D$ → $Y$ represents the influence of the camouflaged edges and features constructed by the bot, which serves to evade detection and interfere with classification.

% According to Figure.\ref{fig:figure_label}, our model performs detection based on $A$ and $D$. 
As stated in the work~\cite{fan2022debiasing}, to quickly reduce the loss function during the optimization process, the model will focus on relying on the substructure with deviations for prediction., i.e., $D$ → $Y$. Therefore, we propose a causal intervention strategy to block the disguised behavior $D$ → $Y$ and reinforce the authentic behavior $A$ → $Y$. In this regard, we simulate different environments and perform causal interventions, generating two heterogeneous social network graphs from the same original graph after causal interventions:
\begin{equation}\label{inter}
P(Y \mid do(A), do(\neg D)).
\end{equation}
where, $do(A)$ represents the intervention on $A$ (reinforcing the authentic behavior). $do(\neg D)$ represents the intervention that blocks the effect of $D$ (removing the spurious influence of $D$ on $Y$.

Specifically, to capture stable associate representations and reduce the influence of biased correlations, we employ distinct environments to simultaneously train a pair of RGCNs, denoted as $G_1$ and $G_2$, based on the original graph $G_o$.

\begin{equation}\label{environ}
G_1 = RGCN_1(G_o), G_2 = RGCN_2(G_o), 
\end{equation}

In this stage, the overall goal is to make the representations of the two RGCNs as far apart as possible. Thus, we define the loss function as:

\begin{equation}\label{equ_kl_loss}
L_{KL} = - D_{KL}(G_1) \parallel (G_2)).
\end{equation}

After that, we use graph structure learning to update $G_1$ and $G_2$:

\begin{equation}\label{equ_rgcn_embedding}
\small{
\begin{split}
x^{(l+1)}_i &= \Theta_{\text{self}} \cdot x^{(l)}_i + \sum_{r \in R} \sum_{j \in N_r^{G_m}(i)} \frac{1}{|N_r^{G_m}(i)|} \Theta_r \cdot x^{(l)}_j,
\end{split}}
\end{equation}
where $G_m$ is the graph in different environments, $\Theta$ is the projection matrix, we convert the user representation via MLP:  

\begin{equation}\label{equ_mlp_embedding}
r_i = \varphi(W_2 \cdot x^{(L)}_i + b_2), 
\end{equation}
where $W_2$ and $b_2$ are learnable parameters, $L$ is the number of layers and $r_i$ is the representation of user $i$.

Finally, we get $r_{k1}$ and $r_{k2}$ using Equation~\eqref{equ_rgcn_embedding} based on $G_1$ and $G_2$. The $G_1$ and $G_2$ are optimized by cross entropy loss, and $L_{KL}$ is used to constrain the distribution of $G_1$ and $G_2$. The overall loss function is:
\begin{equation}\label{equ_inter}
\small{
L_{Inter} = \lambda_1 L_{KL} + (1 - \lambda_1) \Big( L_{CE}(r_{k1}, y) + L_{CE}(r_{k2}, y) \Big),
}
\end{equation}
where $ \lambda_1 $ is a hyperparameter used to balance the impact of the cross entropy loss and the KL loss.

\subsection{Uncertainty Based Bot Detection}
With the evolution of social bots, bots can imitate the characteristics and behaviors of real users, which may lead to false determinations with high confidence. To mitigate this issue in bot detection, our modified uncertainty loss introduces an additive item for false determinations with substantial evidence, aiming at avoiding misclassification due to disguised features of bots.

\textbf{Uncertainty Quantification.} Based on the Dempster-Shaffer Evidence Theory (DST) \cite{yager2008classic}, we use belief quality to analyze model uncertainty \cite{sensoy2018evidential}. Assume that the evidence of each category of samples is represented by $\epsilon$ and the uncertainty of the model output is represented by $U$. Inspired by the definition of uncertainty \cite{sensoy2018evidential}, the uncertainty of bot detection results can be formulated as: 

\begin{equation}
\label{equ_uncertainty}
U = 1 - \sum_{k=1}^{2} \frac{\epsilon_k}{S}, \quad S = \sum_{i=k}^{2} (\epsilon_k + 1),
\end{equation}
where the $S$ represents the normalization factor for the uncertainty distribution in binary classification. 

To model the evidence, we introduce the Dirichlet distribution. The Dirichlet distribution is a probability density function for possible values of the probability mass function $p$, which can be written as:

\begin{equation}
\label{equ_Dirichlet}
D(p|\alpha) = \begin{cases}
\frac{1}{B(\alpha)} \prod_{k=1}^{2} p_k^{\alpha_k - 1}, & \quad p \in \{ p_1,p_2 \} \\
0, & \quad \text{otherwise},
\end{cases}
\end{equation}
%where $\alpha_k$ represents the Dirichlet distribution parameter for classification sample $i$. 
where $\alpha_k$ represents the parameter of the Dirichlet distribution for classifying sample $k$, which is used to map the output distribution of the model to the Gaussian distribution space.

Through these steps, evidence can be inferred from the corresponding Dirichlet distribution parameters, $\epsilon_k = {\alpha_k - 1}$, and $S = \sum_{k=1}^{2} \alpha_k$. Then, substituting $\alpha_k$ into Equation~\eqref{equ_uncertainty} yields the model's uncertainty:

\begin{equation}
\label{equ_uncertainty2}
U = 1 - \sum_{k=1}^{2} \frac{\alpha_k - 1}{\alpha_k} = \frac{2}{S}.
\end{equation}

\textbf{Uncertainty Learning. } As shown in Figure \ref{fig:figure2}, an uncertainty loss is desired to force the uncertainty learning model to learn the parameter $\alpha$ of Dirichlet distribution thus enabling the model to output uncertainty scores. Meanwhile, hard samples (i.e., high-confidence misclassification samples) need to be given extra attention. Therefore, the modified uncertainty loss for bot detection is proposed, which can be formulated as:
% \begin{equation}
% \label{equ_uncertainty_loss}
% \small{
% \begin{aligned}
% L_i (\Theta) 
% &= - \log \left( \int p_{i0}^{y_{i0}} p_{i1}^{y_{i1}} \frac{1}{B(\alpha_i)} p_{i0}^{\alpha_{i0} - 1} p_{i1}^{\alpha_{i1} - 1} \, dp_i \right)\quad \\
%  &\times \left( |Y_i - \hat{Y}_i| \max\left( \frac{1}{u_i}, \delta \right) + 1 \right) \\
% &=\sum_{j=0}^{1}\mathrm{y}_{ij}\left(\log(S_{i})-\log(\alpha_{ij})\right) \\
% &\times
% \begin{cases}
% \max(\frac{1}{u_{i}},\delta) & Y_{i}=\hat{Y_{i}} \\
% 1 & Y_{i}\neq\hat{Y_{i}} & 
% \end{cases}
% \end{aligned}
% }
% \end{equation}
\begin{equation}
\small{
\begin{aligned}
\mathcal{L}_u(\alpha_i)
&=-\lambda_2 \mathrm{log}\left(\int p_{i0}^{y_{i0}}p_{i1}^{y_{i1}}\frac{1}{B(\alpha_i)}p_{i0}^{\alpha_{i0}-1}p_{i1}^{\alpha_{i1}-1}d\mathbf{p}_i\right)\\
&+\left(1-\lambda_2\right)(Y_i-\widehat{Y}_i)^2(1-u_i)\\
&=-\lambda_2\sum_{j=1}^{2} y_{ij} \left( \log(S_i) - \log(\alpha_{ij}) \right)\\
&+\left(1-\lambda_2\right)(Y_i-\widehat{Y}_i)^2(1-u_i),
\end{aligned}
}
\label{Uncertainty_Loss}
\end{equation}
where $y_i$ is a one-hot vector that encodes the true category of the observation $x_i$, $p_{ij}$ is the probability that the $i$-th sample output after model processing belongs to the $j$-th category. The $Y_i$ and $\hat{Y_i}$ are the label and model prediction of $x_i$, respectively, $u_i$ is the uncertainty of $x_i$, and $\lambda_2$ is a hyperparameter used to balance the first part of the loss and the second part of the loss.

We define the optimization loss function of the first part according to the uncertainty formula of the model and design the second part of the loss to enhance the model's learning of difficult samples. Specifically, for the first part of the loss, we adopt the idea of Type \uppercase\expandafter{\romannumeral2} Maximum Likelihood Estimation\cite{seeger2004gaussian} to improve the model's evidence learning ability by optimizing the loss function about the expectation of the Dirichlet distribution ($\alpha_i$). Considering the stealthiness of the bot, we design the second part of the loss to enhance the model’s learning ability for difficult samples, increasing the loss for high-risk nodes with lower uncertainty.

\textbf{Credible Graph Fusion Based on Uncertainty.} After uncertainty learning, we can output uncertainty scores $u_{G_1}$ and $u_{G_2}$ for each specific node (i.e. user) in the graph and fuse a credible graph based on uncertainty scores. Also, through the causal intervention in Section \ref{sec:feature_learning}, we have obtained the graphs $G_1$ and $G_2$ trained in different environments and the corresponding prediction results $\hat{Y}_{G_1}$ and $\hat{Y}_{G_2}$ for each specific node (i.e. user) in the graph. we select the more credible prediction results as the final results $\hat{Y}$ by comparing the uncertainty scores $u_{G_1}$ and $u_{G_2}$, which can be formulated as: % We let the model learn uncertainty reasoning mechanisms from graphs $G_1$ and $G_2$ respectively, so that 
\begin{equation}
\label{equ_u1u2}
\small{
\hat{Y} = h(\hat{Y}_{G_1}, \hat{Y}_{G_2} \mid u_{G_1} ,u_{G_2})=\begin{cases}
\hat{Y}_{G_1}, & u_{G_1} < u_{G_2} \\ 
\hat{Y}_{G_2}, & u_{G_1} \geq u_{G_2},
\end{cases}
}
\end{equation}
where $h(\cdot)$ represents a binary classifier parameterized by $u_{c1}$ and $u_{c2}$, which are the uncertainties of the two causal intervention graphs, and $\hat{Y}$ represents the final prediction.

\section{Experiment}
\subsection{Experiment Settings}
\label{sec:exp_setting}

% \subsubsection{Dataset}
\textbf{Dataset.} We evaluated our method on three widely used social bot detection datasets: Cresci-15 \cite{cresci2015fame}, TwiBot-20  \cite{feng2021twibot}, and TwiBot-22 \cite{feng2022twibot}. Cresci-15 includes 1,950 users, 3,351 bots, 2,827,757 tweets, and 7,086,134 relationships. TwiBot-20 includes 5,237 users, 6,589 bots, 33,488,192 tweets, and 33,716,171 relationships from different fields. TwiBot-22 is the largest open-source Twitter bot detection dataset, covering 860,057 users, 139,943 bots, 86,764,167 tweets, and 170,185,937 relationships. 
For the division of the dataset, we use the dataset's recognized original training, valid, and test splits for fair experimental comparisons.

\begin{table*}[t]
\centering
    \renewcommand{\arraystretch}{0.95}
\begin{tabular}{lcccccc}
\hline
\multicolumn{1}{c}{\multirow{2}{*}{Model}} & \multicolumn{2}{c}{Cresci-15} & \multicolumn{2}{c}{TwiBot-20} & \multicolumn{2}{c}{TwiBot-22} \\ \cline{2-7} 
\multicolumn{1}{c}{}                       & Accuracy      & F1-score      & Accuracy      & F1-score      & Accuracy      & F1-score      \\ \hline
Lee                                         & 98.19 ± 0.07  & 98.52 ± 0.06  & 75.73 ± 0.19  & 79.37 ± 0.19  & -             & -             \\
GAT                                         & 96.44 ± 0.19  & 97.22 ± 0.14  & 77.32 ± 0.73  & 80.51 ± 0.65  & 77.53 ± 0.08  & 53.47 ± 0.46  \\
RoBERTa                                     & 95.70 ± 0.15  & 94.06 ± 0.21  & 74.97 ± 0.23  & 72.80 ± 0.32  & 71.92 ± 0.64  & 16.15 ± 4.98  \\
BotRGCN                                     & 96.37 ± 0.15  & 96.80 ± 0.27  & 83.21 ± 0.37  & 87.68 ± 0.32  & 76.75 ± 0.08  & 48.29 ± 0.66  \\
SATAR                                       & 92.72 ± 0.59  & 93.84 ± 0.52  & 61.70 ± 1.75  & 71.95 ± 0.69  & -             & -             \\
RGT                                         & 96.89 ± 0.16  & 97.58 ± 0.12  & 85.20 ± 0.24  & 86.88 ± 0.22  & \textbf{81.93 ± 0.19}  & 23.85 ± 0.20  \\
BotMoE                                      & 95.30 ± 0.16  & 96.39 ± 0.11  & 84.22 ± 0.34  & 86.89 ± 0.34  & 79.25 ± 0.00   & 56.62± 0.40    \\ \hline
BotUmc                                      & \textbf{98.21 ± 0.11}  & \textbf{98.59 ± 0.09}  & \textbf{87.37 ± 0.06}  & \textbf{89.01 ± 0.05}  & 72.71 ± 0.01  & \textbf{58.52 ± 0.01}  \\ \hline
\end{tabular}
\caption{Accuracy and binary F1 scores of Twitter bot detection methods on three datasets. We run each method five times and report the average value ± the standard deviation. Bold indicates the best performance. Some methods are not scalable to TwiBot-22, indicated by “-”.}
 \label{table:comp}
 \vskip -0.1in
\end{table*}

\textbf{Implement Details.} In the causal intervention stage, we set the learning rate, dropout rate, $\lambda_1$, hidden layer size, and maximum epoch to 1e-2, 0.2, 0.8, 32, and 200, respectively, for the Cresci-15 and TwiBot-20 datasets. For the TwiBot-22 dataset, these parameters are set to 1e-2, 0.2, 0.1, 32, and 3,000. In the uncertainty learning stage, the learning rate, $\lambda_2$, dropout rate, and maximum epoch are set to 5e-5, 0.7, 0, and 100 for the Cresci-15 and TwiBot-20 datasets, respectively. For the TwiBot-22 dataset, these values are set to 1e-5, 0.5, 0, and 50. For fairness, we use the standard public dataset splits to evaluate the methods. For Cresci-15, TwiBot-20, and TwiBot-22, we run them three times on GPU V100, and the average training times of Module 2 are about 55, 90, and 4,148 seconds while the average training times of Module 3 are about 5, 22, and 51 seconds. The used LLM is Llama-3-8b. The numbers of parameters in Module 2 and 3 are 24,036 and 90,214. Baseline methods are described in Appendix \ref{sec:append_baseline}. %; for the three experimental datasets, the convergence time of the model is 1 minute, 2 minutes, and 5 minutes, respectively The performance is evaluated by the F1 score on the test dataset. 

\subsection{Comparative Experiments}
\label{sec:exp_comp}
We compare our proposed BotUmc with 7 representative baselines on three Twitter bot detection benchmarks (Cresci-15 \cite{cresci2015fame}, TwiBot-20 \cite{feng2021twibot} and TwiBot-22 \cite{feng2022twibot}), as shown in Table \ref{table:comp}. Our BotUmc outperforms baseline methods on five out of six results while the Accuracy on TwiBot-22 seems to be poor. However, the F1-score on TwiBot-22 suggests the superiority of our method, which implies the imbalance of samples in the test set. This is supported by the proportions of bot samples in the 3 test sets are 63.2\%, 54.1\%, and 29.4\%, respectively. Therefore, the F1-score can better reflect the performance of the model. In summary, our BotUmc is overall superior to all baseline methods. 

\begin{table}[t]
    \vskip 0in
    \renewcommand{\tabcolsep}{2pt}	
    \renewcommand{\arraystretch}{0.95}
            \begin{center}
               \begin{tabular}{ccccc}
               \hline
Module 1 & Module 2 & Module 3 & Accuracy & F1    \\ \hline
         &          &          & 86.22    & 87.88 \\
         &          & \usym{1F5F8}        & 86.55    & 88.23 \\
         & \usym{1F5F8}       &          & 86.72    & 88.32 \\
\usym{1F5F8}        &          &          & 86.60     & 88.28 \\
         & \usym{1F5F8}        & \usym{1F5F8}        & 86.90     & 88.49 \\
\usym{1F5F8}        & \usym{1F5F8}       &          & 86.81    & 88.41 \\
\usym{1F5F8}        &          & \usym{1F5F8}       & 86.81    & 88.44 \\ \hline
\usym{1F5F8}        & \usym{1F5F8}        & \usym{1F5F8}        & \textbf{87.57}    & \textbf{89.21} \\ \hline
\end{tabular}
    \end{center}
    \vskip -0.1in
    \caption{Ablation study of BotUmc on Twibot-20. Module 1 is the knowledge reasoning module, Module 2 is the interventional graph-based feature learning module, and Module 3 is the uncertainty-based bot detection module. %Bold indicates the best performance.
    }
    \label{table:abl_study}
    \vskip -0.1in
\end{table}

\subsection{Ablation Study}
\label{sec:exp_abl}
Our proposed BotUmc consists of 3 modules: LLM-based Knowledge Reasoning, Interventional Graph-Based Feature Learning, and Uncertainty Based Bot Detection. We perform ablation experiments to show the role of 3 modules, and the results are shown in Table \ref{table:abl_study}.

\textbf{The role of LLM-based Knowledge Reasoning.} We remove the LLM-based knowledge reasoning module, using the original tweet text as the input. The performance degradation implies that the implicit intention mining of tweets and the completion of contextual information are crucial for text representation.

\textbf{The role of interventional graph-based feature learning.} To evaluate the role of interventional graph-based feature learning, we remove the KL loss and train R-GCNs with only cross entropy loss, which in turn leads to a performance degradation shown in Table \ref{table:abl_study}. The degradation suggests that causal interference may find robust associations and improve the learning of stable representations, reducing the impact of Twitter bot disguise behavior on detection.

\textbf{The role of Uncertainty Based Bot Detection.} Here, we remove the Uncertainty Based Bot Detection module to explore the role of uncertainty measurement in social bot detection. Concretely, we select the one with better F1-score from the $G_1$ and $G_2$ output by the interventional graph-based feature learning module as the final results. The performance is still decreased, which shows that the uncertainty-based merging strategy can effectively select more correct detection results. It also implies that the uncertainty training can effectively learn the uncertainty of each user's detection result.%, which is beneficial to the detection of witter bots.

\subsection{Experiment on Hyperparameters}
\label{sec:hyper}
\begin{figure}[t]
		\vskip 0.2in
		\centering
		\subfigure[hyperparameter experiments result of $\lambda_1$]{
			\includegraphics[width=0.75\linewidth]{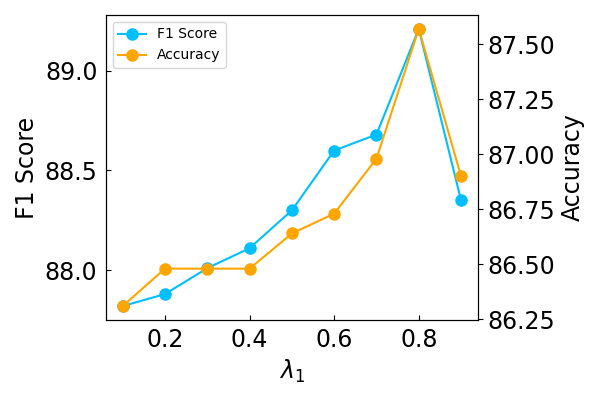}
			\label{fig:hypera}
		}
        
		\subfigure[hyperparameter experiments result of $\lambda_2$]{
			\includegraphics[width=0.75\linewidth]{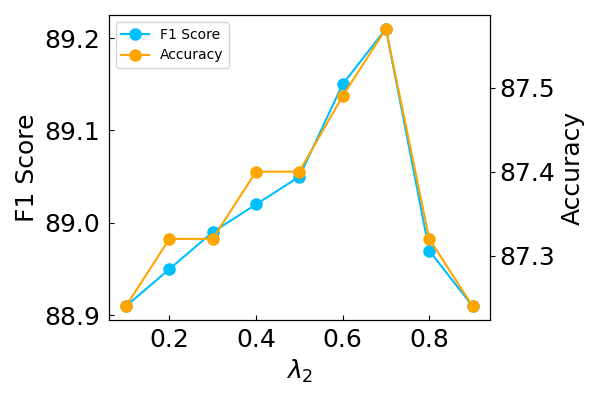}
			\label{fig:hyperb}
		}
		\caption{F1 Scores and Accuracies of our proposed BotUmc with different values of the hyperparameters $\lambda_1$ and $\lambda_2$ on Twibot-20. Both the ranges are 0.1 to 0.9, with an interval of 0.1}
		\label{fig:hyp}
		\vskip -0.1in
	\end{figure}

We conducte experiments on hyperparameters $\lambda_1$ in the Equation~\eqref{equ_inter} and $\lambda_2$ in the Equation~\eqref{Uncertainty_Loss}. The results are shown in Figure \ref{fig:hyp}.

As shown in Figure~\ref{fig:hyp} (a), increasing $\lambda_1$ from 0.1 to 0.8 consistently enhances the model's performance, suggesting that a higher KL loss encourages $G_o$ to learn more robust features across different environments, thereby mitigating the impact of bot disguises. However, when $\lambda_1$ is increased from 0.8 to 0.9, a sharp performance decline is observed. This indicates that excessively high values of $\lambda_1$ lead to an overly large KL loss, causing the model to focus excessively on generating features for diverse environments while neglecting the learning of true value labels.

In Figure~\ref{fig:hyp} (b), as $\lambda_2$ increases, the model gradually prioritizes learning more from the true value labels while minimizing the influence of the false value labels. The model achieves optimal performance when $\lambda_2$ reaches 0.7. However, further increases in $\lambda_2$ lead to a significant performance decline, indicating that an excessively high value of $\lambda_2$ causes the model to focus exclusively on the true value evidence, thereby neglecting the learning from high-risk, difficult samples.

\begin{figure}[t]
    \centering
    \includegraphics[width=1\linewidth]{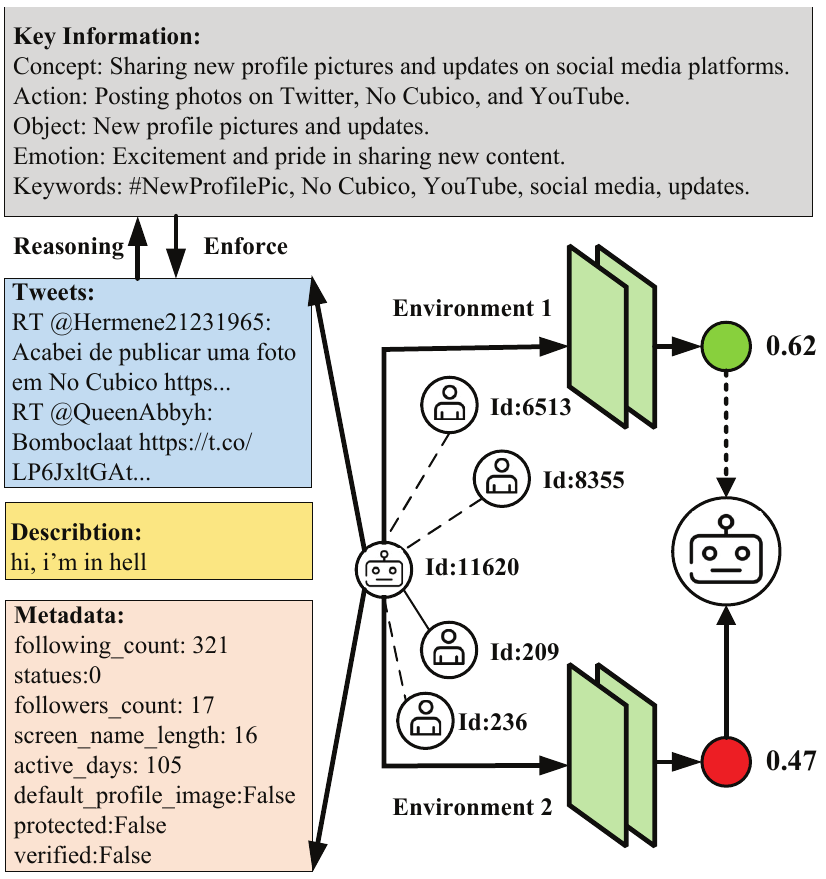}
    \caption{Case study: the text, metadata, and graph information of a hidden bot account, as well as key information extracted from tweets and uncertainty scores of model outputs under different environments.}
    \label{fig:case_study}
    \vskip -0.2in
\end{figure}

\subsection{Case Study}
\label{sec:exp_case}
To demonstrate that our model can effectively identify high-risk disguised bots, we show a typical example in Figure \ref{fig:case_study}. The bot's neighbors are mainly human user accounts, which may cause human features to be aggregated into the current bot features, reducing the probability of the bot being detected. Also, the bot's tweets are deceptive, and it is difficult to distinguish this account from humans'. We use the LLM to effectively extract key information (in the grey box) to enhance the text information. This case shows that the two views provide different predictions. We pick the correct prediction as a final decision by comparing their uncertainty score. 

\section{Conclusion}
We propose a novel uncertainty-aware bot detection method, BotUmc, which quantifies the confidence of its outputs. To leverage uncertainty quantification, we introduce causal perturbations to generate multiple views of social networks in different environments. High-confidence outputs are then selected as the final decisions to enhance performance. Additionally, we design a specialized uncertainty loss function to correct false determinations with high confidence during training, thereby preventing misclassification caused by the subtle features of bots. Compared to existing bot detection methods, BotUmc demonstrates superior performance. 
% In future work, we aim to explore multimodal bot detection approaches to further advance this field.
In future work, we aim to explore multimodal bot detection methods to fully utilize the multimodal information in the social domain to improve the accuracy of bot detection.

\section*{Limitations}
Our work proposes an LLM-based uncertainty-aware Twitter bot detection which can pick the high-confidence prediction from multi-view graphs as the final decision. However, there are still some limitations: 1) Our model does not consider applications on other multimodal tasks, such as video, images, and so on; 2) Since existing bot detection datasets are limited to the Twitter platform and do not cover other social media platforms (such as Facebook, Instagram, etc.). Future research will expand to other social platforms and construct corresponding datasets to further verify the applicability and effectiveness of our method.

\bibliography{custom}

\appendix

\section{Baseline Methods} 
\label{sec:append_baseline}
We compare BotUmc's approach with the following methods: Lee~\cite{lee2011seven} uses random forests with multiple user features to detect bots. GAT~\cite{velivckovic2017graph} uses the graph attention mechanism to adaptively assign weights to node neighbors and capture information in the graph structure for Twitter bot detection. RoBERTa~\cite{liu2019roberta} uses the powerful text representation ability to model user behavior on social media for Twitter bot detection. BotRGCN \cite{feng2021botrgcn} builds a heterogeneous graph of social networks and uses a relational graph convolutional network for Twitter bot detection. SATAR~\cite{feng2021satar} uses semantic, attribute, and neighborhood information for self-supervised learning of Twitter user representations. RGT~\cite{feng2022heterogeneity} effectively learns graph-structured data through the relationships between nodes in the graph for Twitter bot detection. BotMoE~\cite{liu2023botmoe} introduces a community-aware hybrid expert layer to improve the accuracy of Twitter bot detection.

\end{document}